# Near-substrate composition depth profile of direct current-plated and pulse-plated Fe–Ni alloys

Katalin Neuróhr [a,*,1], Attila Csik [b], Kálmán Vad [b], György Molnár [c],
Imre Bakonyi [a], László Péter [a,1]

[a] *Institute for Solid State Physics and Optics, Wigner Research Centre for Physics, Hungarian Academy of Sciences, P.O. Box 49, 1525 Budapest, Hungary*
[b] *Institute of Nuclear Research of the Hungarian Academy of Sciences, P.O. Box 51, 4001 Debrecen, Hungary*
[c] *Research Centre for Natural Sciences, Hungarian Academy of Sciences, P.O. Box 49, 1525 Budapest, Hungary*

ABSTRACT

Composition depth profiles of d.c.-plated and pulse-plated Fe–Ni alloys have been investigated with the reverse depth profile analysis method. When d.c. plating is applied, the mole fraction of iron near the substrate is higher than during steady-state deposition since iron is preferentially deposited beside nickel and the achievement of the steady-state deposition condition takes time. The steady-state composition was achieved typically after depositing a 90-nm-thick alloy layer. In the pulse-plating mode, samples with nearly uniform composition could be obtained at a duty cycle of 0.2 or smaller, and a continuous change in the composition profile could be seen as a function of the duty cycle above this value. A constant sample composition was achieved with pulse-plating in a wide peak current density interval. The composition depth profile was also measured for a wide range of $Fe^{2+}$ concentration. The different characteristics of the composition depth profile as a function of the deposition mode can be explained mostly in terms of mass transport effects. The elucidation of the results is fully in accord with the kinetic models of anomalous codeposition and with the assumption of the superposition of a stationary and a pulsating diffusion layer. The results achieved help to identify the conditions for the deposition of ultrathin magnetic samples with uniform composition along the growth direction.

## 1. Introduction

Electrodeposition is a widely used technique to produce Fe–Ni alloys. The importance of the Fe–Ni alloys stems mainly from their magnetic properties, especially the low coercivity. The first review of this topic was published as early as in 1962 [1].

Fe–Ni plating can be classified as anomalous codeposition [2–7] because the Fe mole fraction in the deposit ($y_{Fe}$) is larger than the corresponding concentration ratio in the solution ($y_{Fe} > c(Fe^{2+})/[c(Fe^{2+}) + c(Ni^{2+})]$), while the order of the thermodynamic nobility of Fe and Ni is opposite. A comparison of the electrodeposition results indicates that the codeposition of binary iron-group alloys is not totally analogous. It was found that codeposition of Ni–Co and Fe–Ni shows more mass-transfer effects than does Co–Fe deposition within the current density ranges studied [8].

Regardless of the specific solution components, the mole fraction of iron increases monotonically with the $Fe^{2+}$:$Ni^{2+}$ solution concentration ratio [3,9–11]. The study of the codeposition kinetics of Fe and Ni showed that nickel reduction is inhibited while iron deposition rate is enhanced when compared with the deposition rates of the individual metals in single-metal plating baths [5,12].

The effect of various electrolyte components on Fe–Ni deposition has also been investigated. Boric acid prevents the passivation of the electrode surface [13,14], increases the rate of Fe deposition during the deposition of Fe–Ni alloys [15,16] and reduces the oxygen inclusion into the deposit [17]. The presence of citric acid in the plating solution increased the iron content of the deposit, while the presence of ascorbic acid had no important effect [16]. In contrast, tartaric acid and ethylenediamine decreased the Fe content of the deposits [18]. The inhibition effects of nickel and iron ions on proton reduction were demonstrated in chloride, sulfate and perchlorate solutions [7]. It was also shown that the concentration of sulfur inclusion plays a crucial role in the corrosion rate of electrodeposited Fe–Ni alloys [19].

Fe–Ni deposits containing either a Fe-rich body-centered cubic or a Ni-rich face-centered cubic crystal structure can form. Generally, a 10 wt.% wide composition range occurs in which both these crystalline forms can co-exist in what is called a "duplex structure" [3,11,20–22]. The achievement of nanocrystalline deposits is a key issue nowadays, but very divergent trends have been observed for the grain size as a function of the deposit composition [3,9,11,23–25].

* Corresponding author. Tel.: +36 1 392 2222; fax: +36 1 392 2215.
  *E-mail address:* neurohr.katalin@wigner.mta.hu (K. Neuróhr).
[1] ISE member.

Several attempts have been made to describe the kinetics of the deposition of binary and ternary alloys of the iron-group metals under potentiostatic conditions. The common features of all models proposed so far are that a competitive adsorption of the species containing the metal ions and a one-dimensional diffusion process are considered. One group of the models disregards the two-step reduction process [12,26–29], but more complex descriptions take into account the formation of the Me(I) intermediates [30,31] and, furthermore, the interrelation of the metal ion reduction processes in some autocatalytic reactions [4]. In some cases, a special goal of the kinetic modeling was the evaluation of the surface pH and the determination of the role of the hydrogen ions in the metal reduction process [30,32]. In all above mentioned studies, the calculations have been made for steady-state deposition conditions, disregarding all possible transient effects. Since all these models are fairly complicated, no explicit equations could be given for either the current density or the deposit composition as a function of the electrode potential, but rather a particular set of experimental conditions could be successfully modeled by the kinetic scheme proposed.

Pulsed electrodeposition of Fe–Ni alloys was studied by several groups with the motivation of studying the nucleation process [33], the investigation of the impact of hydrodynamic conditions [21], the optimization of nanomechanical [24], morphological [15] or magnetic properties [34], and the achievement of an even lateral component distribution [35]. However, the impact of the pulse deposition mode on the composition depth profile received no attention.

The average composition of electrodeposited Fe–Ni films depends on the thickness of the Fe–Ni film. It was shown first by Cockett and Spencer-Timms [36] that the average composition of electrodeposited Fe–Ni alloys depends on the deposit thickness. In the latter work, d.c.-plating and pulse-plating was also compared, but the difference in the initial and steady-state composition was not diminished significantly by the pulsed deposition mode, although pulse-plated samples were more iron-rich than the d.c.-plated ones. This could be due to the large on-time used (1 s [36]), which provides deposition conditions similar to the d.c. deposition. Doyle also showed that the Ni content of the Fe–Ni alloys changed quite abruptly in the first 100 nm and it became stable at about 150–200 nm [37]. Later, Lommel and Girard obtained similar results with three commercial Fe–Ni plating baths [38]. The problem of the composition gradient was later addressed by Beltowska-Lehman and Riesenkampf [39]. They found that the initial composition gradient can be somewhat suppressed if pulse-plating is applied. All the above studies relied on the determination of the average composition of the deposits of various thicknesses; therefore, the exact composition depth profile function could be estimated only rather than being directly measured. The overall composition measurement as an indirect depth profiling method is used even nowadays to measure the thickness-dependent composition of electrodeposited samples [40].

The composition gradient in electrodeposited Fe–Ni alloys was directly observed by Gao et al. [41]. In the latter study, the Auger electron spectroscopic composition depth profile of an electrodeposited Fe–Ni film showed a slight Fe accumulation in the near-substrate zone. The authors attributed this uneven composition near the Si substrate to the preferential deposition of Fe in the early stage of nucleation and growth, but the possible mass transport effects were not mentioned [41].

The accumulation of the preferentially deposited metal in the near-substrate zone seems to be a general phenomenon, as shown by earlier studies of the present authors. In stagnant solutions, the thickness zone needed for the composition stabilization was 100–150 nm for various electrodeposited alloys (Fe–Co–Ni, Cu–Co–Ni, Cd–Ni, etc.) [42–44]. A sensitive analysis of the near-substrate composition evolution was made possible by the reverse composition depth profile analysis, in which the adlayer-supported samples can be separated from the substrate and the sputtering can be started from the region where the deposition was started.

In the present study, the composition depth profile of electrodeposited Fe–Ni samples was investigated by using secondary neutral mass spectrometry (SNMS). The near-substrate composition evolution of samples produced with d.c. plating and pulse plating was analyzed with the reverse composition depth profile analysis method. It was found that, under optimized pulse conditions, it is possible to eliminate the composition variation in the vicinity of the substrate.

## 2. Experimental

### 2.1. Chemicals and materials

The substrate was a metal-coated Si wafer with (1 0 0) orientation. The deposition of a 5 nm thick Cr adhesive layer was followed by that of a 20 nm thick Cu conducting layer. Both the Cr and Cu layers were produced by evaporation. The mean surface roughness of the Si/Cr/Cu substrates was determined with atomic force microscopy. The root-mean-square surface roughness of the substrate was found to be between 1 and 3 nm [45].

All chemicals used for the solution preparation were of analytical grade, except for the $NiSO_4$ used for depositing the supporting Ni adlayer. The solutions were prepared with ultrapure water (resistivity: 18 MΩ cm). Electrolytes containing $Fe^{2+}$ ions were freshly prepared every day in order to avoid the formation of $Fe^{3+}$. Fe–Ni samples were made from the following electrolyte: ultrapure $NiSO_4$ (0.55 mol dm$^{-3}$ unless otherwise mentioned), $FeSO_4$ (0.045 mol dm$^{-3}$ unless otherwise mentioned), $Na_2SO_4$ (0.3 mol dm$^{-3}$), $H_3BO_3$ (0.1 mol dm$^{-3}$), saccharin (0.2 g dm$^{-3}$) and sodium dodecylsulfate (0.03 g dm$^{-3}$). The electrolyte pH was set to 2.8 by adding $H_2SO_4$ to the solution. Boric acid served as a buffering agent to retard metal hydroxide precipitation [14], while saccharin and sodium dodecylsulfate were added in order to promote the deposition of a smooth coating without any pits forming due to hydrogen evolution.

The Ni adlayer was plated from a solution composed of technical grade $NiSO_4$ (0.60 mol dm$^{-3}$), $Na_2SO_4$ (0.20 mol dm$^{-3}$), $MgSO_4$ (0.16 mol dm$^{-3}$), $NaCl$ (0.12 mol dm$^{-3}$) and $H_3BO_3$ (0.40 mol dm$^{-3}$). This composition was chosen to ensure that a sufficiently tensile Ni support was deposited onto the Fe–Ni layers with low internal stress. Advantage was taken of the presence of Co in the technical grade nickel sulfate to detect the transition region between the Fe–Ni layer and the Ni cover layer.

### 2.2. Electrodeposition

Electrodeposition was carried out in a tubular cell. The exposed surface area of the upward facing cathode was about 8 mm × 20 mm, and the recessed part of the cell was 15 mm high, hence ensuring an even accessibility of the entire cathode surface. The counter electrode was a Ni sheet immersed parallel to the cathode at the top of the cell. For the preliminary cyclic voltammetric study and for the chronopotentiometric measurements, a Pt or Cu working electrode and a saturated calomel reference electrode (SCE) was used.

Electrodeposition of the samples meant for the depth profile analysis was carried out galvanostatically on the Si/Cr/Cu substrates using a 2-electrode configuration. Once the Fe–Ni layer was formed, it was followed by the deposition of a Ni support layer on the top. The preparation of the subsequent electrodeposited layers was performed by changing the electrolytes but without disassembling the cell. This method assured that the same area was covered completely with the subsequent layers. The minimum total thickness of the supporting layer was about 3 μm in order to achieve a sufficient toughness that enabled us to peel off the deposits from the substrate without any significant damage. Further details of the sample preparation process can be found in the earlier papers [42–45].

Samples were obtained under galvanostatic control by either d.c. plating or pulse plating with an EF453 type potentiostat/galvanostat (Electroflex, Hungary). In the case of pulse plating, the total cycle time was 0.5 s with a typical duty cycle of $\varepsilon = t_{ON}/(t_{ON} + t_{OFF}) = 0.2$, unless otherwise mentioned. The smallest duty cycle applied was 0.04.

After the electrochemical sample preparation, the Si wafer was broken behind the sample in a manner that the deposit itself remained intact. Then, the deposit could be easily peeled off from the Si wafer so that the separation took place at the Si/Cr interface. The surface of the resulting Cr-capped sample was as smooth as the Si wafer, making it particularly appropriate for a sensitive depth profile study.

### 2.3. SNMS measurements and calculation of the composition depth profile functions

The depth profile analysis of the samples was performed by secondary neutral mass spectrometry (SNMS) with an instrument of INA-X type (SPECS GmbH, Berlin, Germany) in the direct bombardment mode by using Ar$^+$ ions with a fairly low energy for sputtering ($E_{Ar^+} = 350$ eV). The erosion area was confined to a circle of 2 mm in diameter by means of a Ta mask. The lateral homogeneity of the ion bombardment was checked by profilometric analysis of the craters sputtered. The method of the determination of the sputtering rates was described earlier ([46] and references cited therein). The uncertainty in the sputtering rate due to the varying composition is lower than 5% of the average value.

## 3. Results

### 3.1. Polarization characteristics of the Fe–Ni system

The potential dependence of the deposition of Fe, Ni and Fe–Ni alloys was studied in comparison with a blank solution. The blank solution contained MgSO₄ instead of the iron-group metal sulfates at the same concentration as the original component (NiSO₄ or FeSO₄) of the plating bath. In all solutions, the occurrence of a peak at −0.41 V can be associated with the onset of the reduction of the H₃O$^+$ ions. Due to the low H$^+$ concentration and the presence of other components in the electrolyte (e.g., saccharin and sodium dodecylsulfate), neither the hydrogen adsorption peaks nor the corresponding hydrogen oxidation peaks can be identified even on a Pt electrode. In the presence of the Ni$^{2+}$ and Fe$^{2+}$ cations, the H₃O$^+$ reduction peak can still be seen, and the stripping behavior of each metal can also be established. The difference in the cathodic current densities for the Ni and Fe–Ni electrolytes shows that iron has an inhibitory effect for the Ni deposition. The stripping of Ni takes place in a wide potential range because of the kinetic hindrance of Ni dissolution, and a similar behavior is observed for the Fe–Ni alloy. The polarization behavior of the Fe–Ni electrolyte is in accord with the literature data in the sense that the deposition regime of these metals cannot be separated when both metal ions are present [14,15,20,32–34,41,47,48]. At potentials more negative than −1 V, a monotonous increase in cathodic current density is observed.

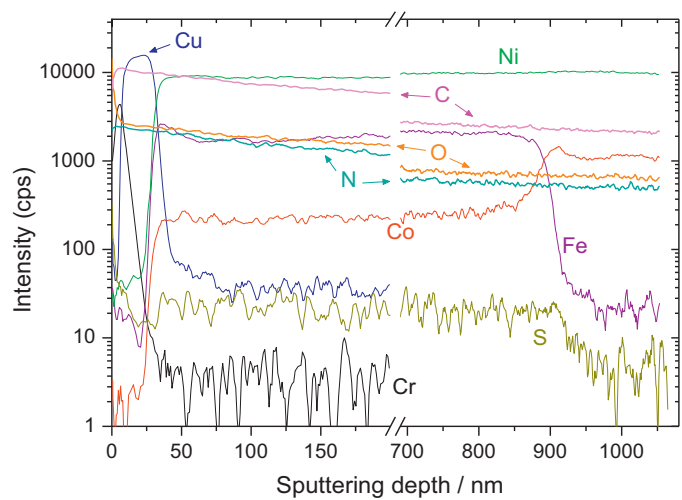

**Fig. 1.** Result of a typical SNMS measurement showing the signal intensity of all important components and impurities. The Cr and Cu signals are cut off after the axis break for sake of clarity.

Hence, the polarization curves do not yield any information on the deposit composition.

### 3.2. General characteristics of the composition depth profiles measured

Fig. 1 presents the composition depth profile of a number of elements in terms of the SNMS signal intensities measured. The common features of the SNMS spectra obtained are as follows.

Clear correlations can be seen in the change of the mole fraction of the main elements once a layer boundary is reached with the sputtering front. When the sputtering of the Cu layer is finished and the sputtering of the Fe–Ni layer is started, the loss of the Cu signal is accompanied by the rise of both the Fe and Ni intensity. At the Fe–Ni/Ni(Co) boundary, the decrease in the Fe and the increase of the Co count intensity are also accompanied with a step in the sulfur signal intensity, which clearly indicates that sulfur in the Fe–Ni layer originates from the saccharin (this component was not present in the solution used for depositing the Ni cover layer). As it can be seen in the intensity vs. sputtering time function, the Ni signal does not change when the Fe–Ni//Ni(Co) boundary is crossed with the sputtering front. The reason of the apparently invariant Ni signal stems from the small difference in the Ni content of the two neighboring layers in the actual sample. The larger the Fe content of the Fe–Ni layer, the larger step can be seen in the Ni signal at this boundary, too (although the logarithmic scale applied to present all signals in a single figure tends to hide this step).

The determination of some impurities was not possible with the method applied for the composition depth profile analysis. Light elements such as carbon, nitrogen and oxygen are relatively abundant components in the residual gas of the high-vacuum SNMS instrument. Since the great majority of the O, C and N intensity comes from the residual gases, their concentration in the sample being sputtered cannot be determined at the trace level due to the time-dependence of the background signal. The absence of non-metallic inclusions can be inferred from the observation that the signal intensity for C, N and O do not undergo a stepwise change when the sputtering reaches the Fe–Ni//Ni boundary. It is also likely that these elements originate from a single source since their signal intensities change in parallel throughout the entire sputtering process. However, deposits made of the iron group metals always have a low non-metallic content, the upper limit of which is assessed from the data measured in the present study to be 0.2 at.%. Due to the uncertainty of the background for these light elements, their mole fraction was omitted from the composition depth profiles shown in the following sections.

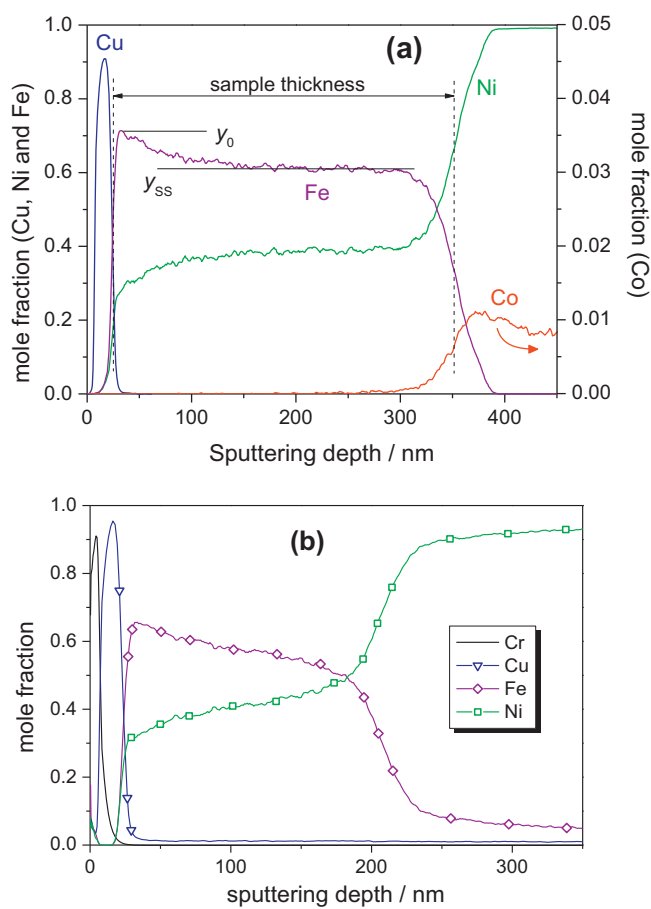

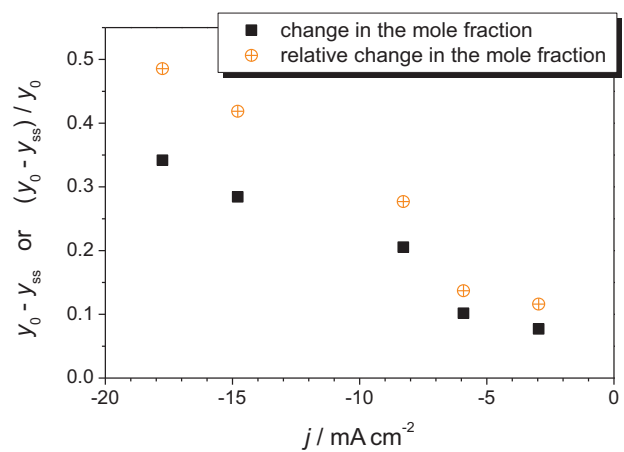

**Fig. 3.** The absolute and relative difference of the initial and steady-state Fe mole fraction of the deposits as a function of current density of the d.c.-plated samples.

**Fig. 2.** Representative reverse composition depth profiles of two d.c.-plated Fe–Ni samples. Deposition conditions: (a) $-5.9$ mA/cm$^2$, deposition time: 250 s; average current efficiency: 66%; here, the mole fraction of Cr is omitted for sake of clarity. (b) $-2.96$ mA/cm$^2$, deposition time: 504 s; average current efficiency: 35%. In both cases, $c(Fe^{2+}) = 0.045$ mol dm$^{-3}$.

Our observation concerning the oxygen content of the deposits is in agreement with the findings of Gadad and Harris [17] in the sense that solutions containing boric acid lead to deposits with low oxygen content. However, Tabakovic et al. [49] found that electrolytes prepared from the chlorides of the metals lead to a deposit with several percent of oxygen. This draws the attention to the importance of the type of metal salts used in the plating process.

### 3.3. Composition depth profile of the d.c.-plated samples

Fig. 2 shows the reverse composition depth profile functions of two d.c.-plated Fe–Ni samples. Fig. 2a presents the establishment of the key quantities to be used for the characterization of the samples. The base of a Fe–Ni deposit is set by the sputtering depth ($d_{SP}$) at which the Cu layer ends and the Fe content of the sample rises. The top of the Fe–Ni layer is defined by the inflection point of the depth profiles of the three elements Ni, Fe and Co present in the technical grade nickel sulfate used to deposit the Ni support layer. The average sample thickness is then determined from the distance between the base and top of the Ni–Fe layer. Fig. 2a also shows the initial and steady-state mole fractions of Fe ($y_0$ and $y_{SS}$, respectively). The Fe mole fraction at the innermost location is determined by extrapolation of the Fe profile to the Cu//Fe–Ni boundary.

The steady-state mole fractions of the d.c.-plated deposits could be easily obtained from the composition depth profile functions when the deposition current density was higher than approximately $-6$ mA cm$^{-2}$. In these samples, the Fe mole fraction decreases around the Fe–Ni//Ni(Co) interface over a distance less than 100 nm thick. The mean deposit thickness can be determined from the sputtering depth at which the mole fraction vs. depth function has an inflection point (as shown in Fig. 2a). However, at low current densities, the current efficiency is so low that the sample thickness is much below the nominal value (i.e., based on 100% current efficiency) and the region over which the Fe composition declines extends to more than 200 nm. In parallel with the decrease in current efficiency, the surface roughness of the samples also increased, which could be seen even by visual observation. Therefore, the Fe–Ni//Ni(Co) boundary was much less sharp than in the ideal cases. All these features can be seen in the composition depth profile function shown in Fig. 2b.

The dependence of the uniformity of the composition of the d.c.-plated samples on the current density is summarized in Fig. 3. The non-uniformity of the deposits in terms of the change in the mole fractions increases almost linearly with the current density for the electrolyte used. At current densities higher than $-15$ mA cm$^{-2}$, the steady-state mole fraction of Fe is about half of the initial value. Fig. 3 clearly shows that it is not possible to produce ultrathin deposits with uniform composition by d.c. plating.

Chronopotentiometric experiments have also been performed to determine if the composition modulation in the near-substrate zone is accompanied by a simultaneous change in the electrode potential of the cathode. These data (not shown) showed an abrupt change in electrode potential at the beginning of the deposition, partly due to a capacitive transient. The initial spike decayed over 5–6 s, and the length of this transient period was not a function of the current density in the $-3$ to $-35$ mA cm$^{-2}$ current density range. In all cases, the steady-state deposition potential was 60–70 mV less negative than the initial potential peak. Therefore, it is concluded that this potential transient at the beginning of the deposition is not directly related to the composition change in the deposit. Since the thickness over which the deposit composition varies does not depend on the current density, the transient period should become shorter as the applied current increases. However, this trend cannot be observed in the chronopotentiometric curves; instead, a constant potential decay period was observed.

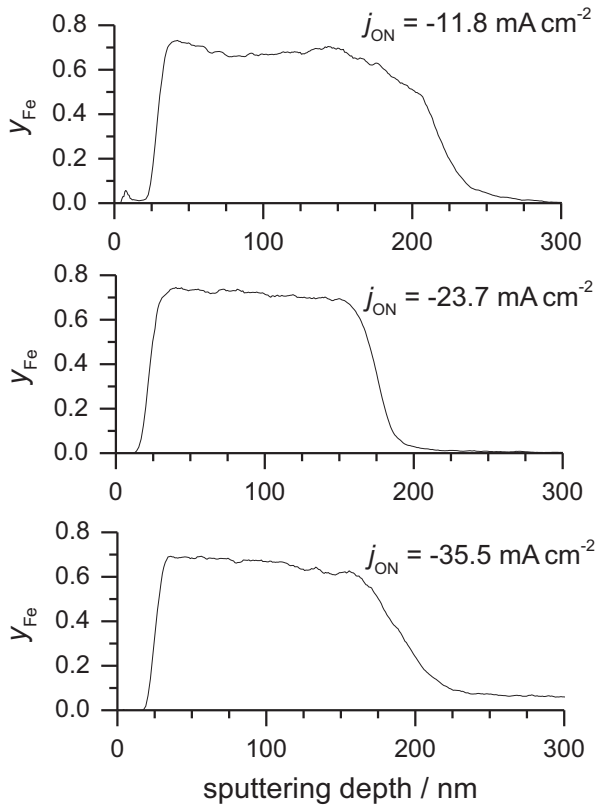

**Fig. 4.** Composition depth profile functions of three pulse-plated Fe–Ni samples (for sake of clarity, only the mole fraction of Fe is shown). The current density during the pulse time is indicated next to each curve. $t_{ON} = 0.1$ s, $t_{OFF} = 0.4$ s, $c(Fe^{2+}) = 0.045$ mol dm$^{-3}$.

### 3.4. Composition depth profile of the pulse-plated samples

Fig. 4 shows the reverse composition depth profile functions of three pulse-plated Fe–Ni samples (for the sake of clarity, only the mole fraction of Fe is shown). The current density during the on-time is indicated next to each curve. In the case of the pulse-plated samples, the initial and steady-state Fe mole fraction differ to a negligible extent only. The $t_{OFF} = 0.4$ s relaxation time was long enough to recover the electrolyte concentrations near the cathode and, hence, the next pulse could lead to a deposit of the same composition as during the previous pulse.

A much wider range of current densities can be applied during pulse plating than during d.c. plating. While d.c.-plated samples obtained with current densities larger than $-30$ mA cm$^{-2}$ were visually very rough and porous, the pulse-plating method enabled us to use as high as $-85$ mA cm$^{-2}$ peak current density at the 0.2 duty cycle chosen. The applicability of this high current density is in accord with the general theory of pulse plating [50,51]. Even at the $-85$ mA cm$^{-2}$ peak current density, a homogeneous composition depth profile could be measured.

Fig. 5a summarizes the initial and steady-state compositions of both d.c.-plated and pulse-plated Fe–Ni samples. As also shown in Fig. 3, the composition of the d.c.-plated samples varies after the start of the deposition. The comparison of the initial and steady-state composition of the pulse-plated samples indicates, however, that the composition change in the initial stage of the deposition is very small. Additionally, since the composition changes appear to be random fluctuations and show no discernible dependence on the current density, they can be attributed to experimental errors. The steady-state composition of the pulse-plated samples was independent of the current density over the range from $-12$ to $-30$ mA cm$^{-2}$. Interestingly, the current density range of the invariant deposit composition coincides with that where the current efficiency is high, as shown in Fig. 5b. The current efficiency was calculated as the ratio of the sample thickness derived from the composition depth profile function (according to Fig. 2a) and that calculated from Faraday's law by assuming 100% current efficiency. Although the data are somewhat scattered, a pronounced current efficiency maximum can be seen at $-20$ mA cm$^{-2}$.

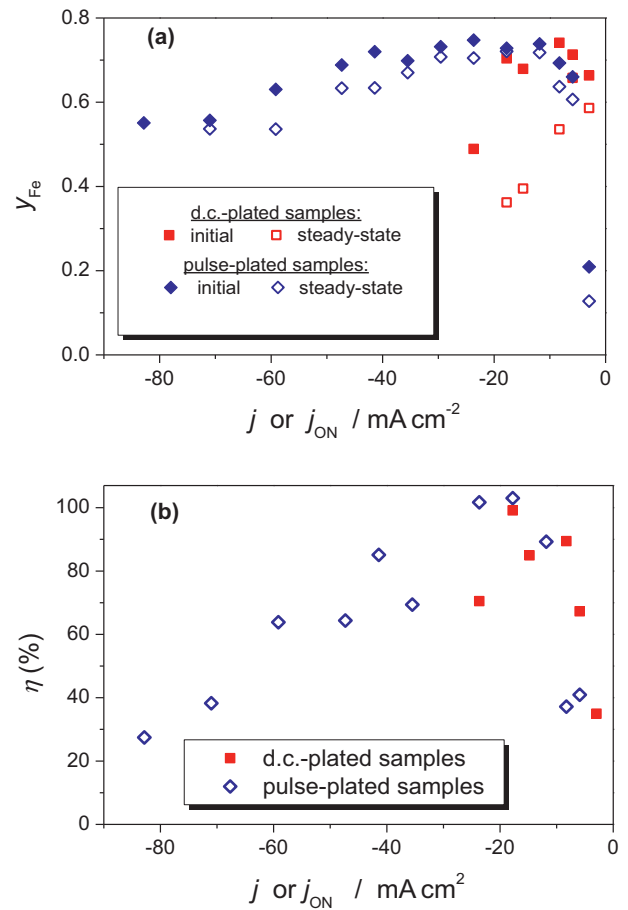

**Fig. 5.** (a) The initial and steady-state compositions of both d.c.-plated and pulse-plated samples as a function of the current density. For pulse-plated samples, current density during the on-time is indicated; duty cycle: 0.2, $c(Fe^{2+}) = 0.045$ mol dm$^{-3}$. (b) The current efficiency of the deposition as a function of the current density, for the same samples as shown in (a).

The method used for the calculation of the current efficiency is based on the fact that the samples comprising both the Fe–Ni and the Ni(Co) layers are dense and contain no cavity; in other words, the samples studied with SNMS exhibit no internal porosity. The electrolyte for depositing the supporting layer was selected to provide a conformal coverage of the layer of interest. If any fluid inclusion (gas bubble or residue of the electrolyte solution) could be present, the SNMS signal intensities would indicate it sensitively. If a non-solid inclusion is reached during the sputtering process, its content is readily atomized in the Ar plasma. This would also lead to a sudden positive change both in the ionization yield and in the detection intensity of the light elements, particularly that of oxygen. Since such effect has not been observed at all, we can be sure of that the entire deposit comprising all layers is free of any cavity. This observation is taken as a validation of the thickness estimation method based on the depth profile analysis.

Fig. 6 shows the comparison of a d.c.-plated and a pulse-plated sample in which the current density was identical ($j_{DC} = j_{ON}$). The

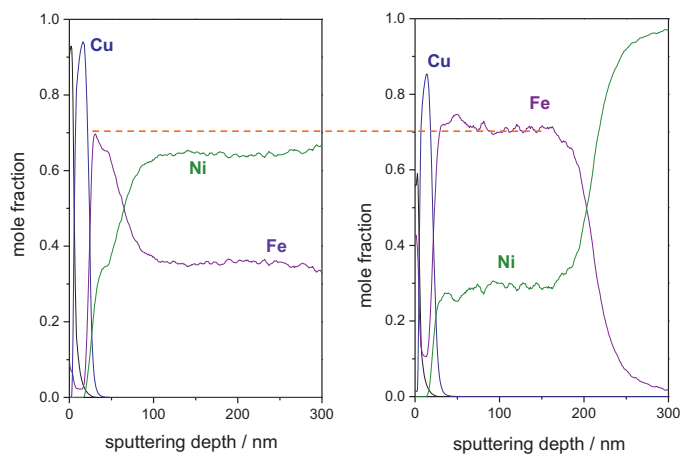

**Fig. 6.** Reverse composition depth profile of a d.c.-plated (left) and a pulse-plated (right) Fe–Ni sample. Current density: $j_{DC} = j_{ON} = -17.75\,\text{mA cm}^{-2}$; $c(\text{Fe}^{2+}) = 0.045\,\text{mol dm}^{-3}$. For the pulse-plated sample, $\varepsilon = 0.2$. The dashed line serves as a guide for the eye to visualize the equality of the initial and steady-state mole fraction in the d.c.-plated and pulse-plated sample, respectively.

figure clearly shows that the initial composition of the d.c.-plated sample is conserved throughout the entire sample thickness when the deposition was carried out with pulse plating. Due to the relaxation of the electrolyte concentration during the off time, each pulse leads to the deposition of an alloy of identical composition. Thus, the duty cycle should be decreased in order to reduce the composition variation across the deposit thickness.

Fig. 7 shows the thickness dependence of the local composition of a number of samples deposited with varying duty cycle. In Fig. 7, the near-substrate zone is magnified and the substrate thickness was deducted from the sputtering depth in order to show all data on an identical depth scale. It can be seen very well that the initial mole fraction of Fe is about $0.71 \pm 0.03$ for each sample. The larger the duty cycle, the higher is the deviation from the initial value. At $\varepsilon = 0.12$ and 0.2, the mole fraction of Fe in the sample seems to be constant along the entire sample thickness. In the case of the d.c.-plated sample (i.e., $\varepsilon = 1$), the steady-state composition is achieved after a deposit thickness of 90 nm and the mole fraction of Fe is just half of the initial value for the composition stabilization.

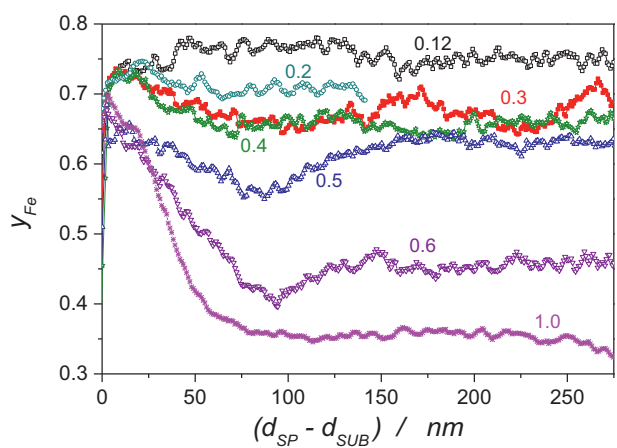

**Fig. 7.** Reverse composition depth profile function of Fe for samples deposited at varying duty cycle (indicated next to each curve). The abscissa shows the sputtering thickness ($d_{SP}$) corrected with the substrate thickness ($d_{SUB}$). Current density: $j_{ON} = -17.75\,\text{mA cm}^{-2}$; $c(\text{Fe}^{2+}) = 0.045\,\text{mol dm}^{-3}$.

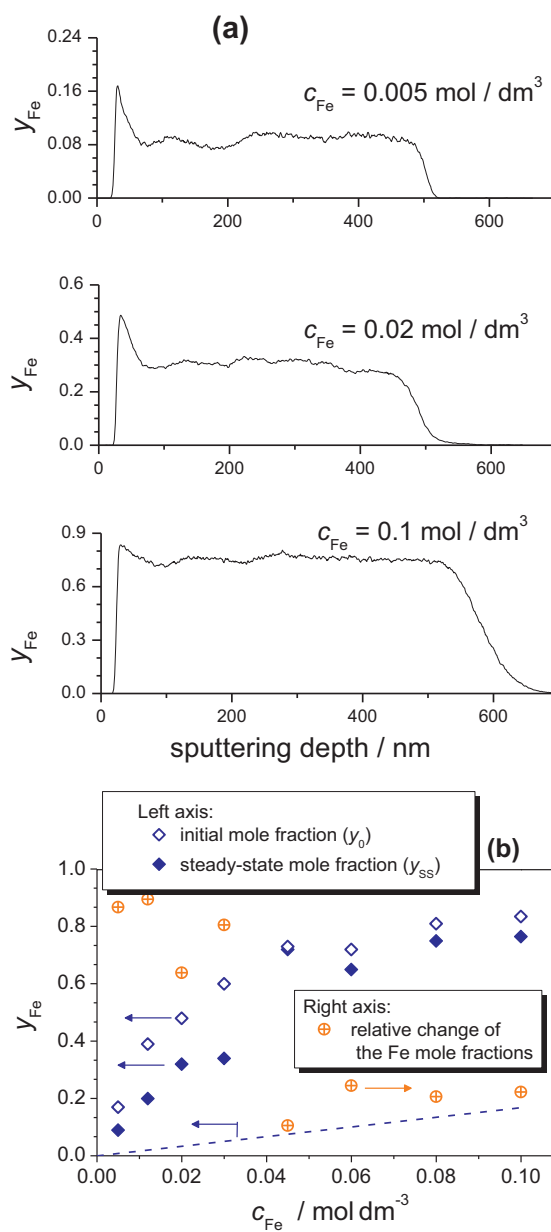

**Fig. 8.** (a) Reverse composition depth profile curves of three samples deposited from solutions of various $\text{Fe}^{2+}$ concentrations, as indicated above the measurement data. (b) Initial and steady-state mole fraction of Fe for pulse-plated deposits obtained with different $\text{Fe}^{2+}$ concentrations. Current density: $j_{ON} = -17.75\,\text{mA cm}^{-2}$; duty cycle: 0.2. The dashed line represents the so-called reference line: $y = c(\text{Fe}^{2+})/[c(\text{Fe}^{2+}) + c(\text{Ni}^{2+})]$.

### 3.5. Impact of the $Fe^{2+}$ concentration

Fig. 8a shows a few reverse composition depth profile curves as a function of the $\text{Fe}^{2+}$ concentration in the bath and Fig. 8b presents the initial and steady-state compositions of pulse-plated samples produced with various $\text{Fe}^{2+}$ concentrations. Other parameters such as the duty cycle, the current density and the overall metal ion concentration were kept constant. As can be expected, both the initial and steady-state Fe concentrations increase with $c(\text{Fe}^{2+})$. For the experimental conditions applied here, the composition of the samples does not change significantly with the thickness if $c(\text{Fe}^{2+}) > 0.1\,\text{mol dm}^{-3}$.

The relative change in the mole fractions $(y_0 - y_{SS})/y_0$, is indicative of the partial current density of Fe deposition. This ratio is between 0.7 and 0.8 if $c(Fe^{2+}) < 0.03$ mol dm$^{-3}$ and decreases to below 0.1 for higher concentrations. One possible explanation for this behavior is that the initial and steady-state compositions should tend toward each other when the steady-state composition is no longer a function of the dissolved $Fe^{2+}$ concentration. In this concentration range, the electrolyte layer near the cathode becomes less and less depleted of the $Fe^{2+}$ ions so that the change in conditions at the beginning and end of each pulse cycle becomes progressively smaller.

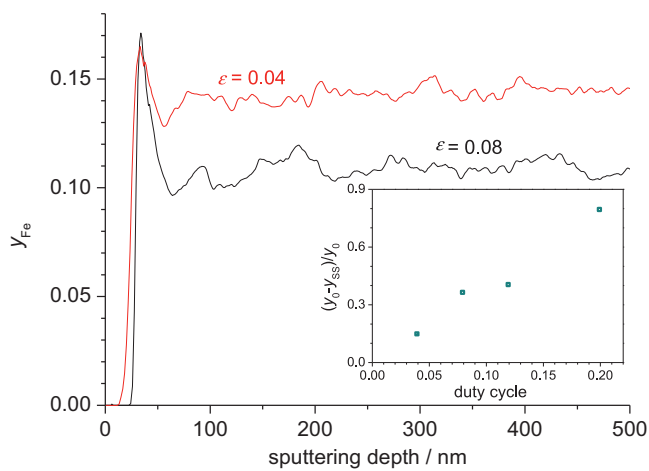

**Fig. 9.** (a) Reverse composition depth profile curves of two samples deposited from solutions of low $Fe^{2+}$ concentration (5 mM) with different duty cycles, as indicated next to the curves. Inset: the relative change of the Fe mole fraction as a function of the duty cycle for samples deposited from a solution of $c(Fe^{2+}) = 5$ mM. Current density: $j_{ON} = -17.75$ mA cm$^{-2}$.

### 3.6. Impact of the duty cycle at low $Fe^{2+}$ concentrations

The dependence of the Fe mole fraction change on the $Fe^{2+}$ concentration at a fixed duty cycle of 0.2 shows that the relative importance of the activation and diffusion control of the Fe deposition is a decisive factor in the formation of the composition gradient. It is expected that at small $Fe^{2+}$ concentrations, the Fe deposition is mass transport controlled, and the full recovery of the initial deposition conditions (i.e., $Fe^{2+}$ surface concentration) is crucial. This would mean that the duty cycle should be decreased in order to eliminate the initially formed mole fraction change in the deposit.

Fig. 9 shows two composition depth profile functions measured for samples prepared with $c(Fe^{2+}) = 0.005$ mol dm$^{-3}$ and small duty cycles. These experiments verified the expectation that the smaller the $Fe^{2+}$ concentration, the lower must be the duty cycle to maintain the variation in deposit composition below a desired level. The inset of Fig. 9 allows us to conclude that the variation in deposit composition can be decreased to nearly zero by extrapolating the duty cycle to $\varepsilon = 0$ if the cathode reaction of the preferentially deposited alloy component is diffusion-controlled.

### 4. Discussion

It has been shown with the present experiments that the Fe–Ni alloys deposited with pulse plating exhibit a much smaller spontaneous composition modulation along the growth direction than their d.c.-plated counterparts. This can be fully explained with the general concept of pulse plating. The smaller the duty cycle, the larger the extent of the concentration relaxation of the reactants near the cathode surface. Therefore, the decrease in duty cycle helps to maintain identical deposition conditions at the beginning of each pulse. At a sufficiently small duty cycle (which is also a function of both the pulse current density and the ion concentrations), the composition profile of the pulse-plated samples becomes nearly flat and identical with the initial composition of the d.c.-plated samples if the pulse current and the d.c. currents are the same. This is well demonstrated in Fig. 6.

The deposit thickness achieved over a typical pulse length of 0.1 s is between 0.2 and 1.0 nm, depending on the current density and the current efficiency. As can be seen in the near-substrate zone of the d.c. deposits, a composition gradient may develop along the growth direction also within a pulse time of 0.1 s. It is obvious that the composition change of the deposit within such a small thickness range cannot be observed for various reasons. The deposit thickness accumulated during a single pulse is comparable to the resolution of the depth profiling method because the time required to make one measurement of the SNMS (i.e., to count the intensity of each element) is of the same order of magnitude as the time required to sputter a 1-nm thick deposit layer. It has to be taken into account that the deposit surface is not completely flat, and the growth of the deposit does not take place with a layer-by-layer mechanism at the atomic scale. Therefore, it is not possible to detect with a depth profiling method whether there is a change in the deposition rate of the alloy components during a single pulse.

It is also believed that the composition change over a thickness of 1 nm is insignificant in all practical applications. The composition change leads to a stress in the deposit due to the varying average nearest neighbor distance. This stress is known to be harmful for the soft magnetic properties of the Fe–Ni alloys by increasing the coercive field. However, the length scale of the exchange interaction between the conduction electrons is much larger than the deposit thickness accumulated during a single pulse, and hence, small local fluctuations in the deposit composition tend to level off.

As shown in Figs. 6 and 7, the deposit thickness necessary for the composition stabilization is about 90 nm for the d.c.-plated samples. It is difficult to compare this value with those reported in the literature since the previous studies relying on an analysis of the total deposit composition only measured the integral of the $y(d)$ function and hence, the thickness of the composition stabilization was uncertain. As can be estimated from the early studies of the field [36–39], the thickness of composition stabilization is 90–250 nm for the d.c.-plated samples and less than 90 nm for pulse-plated samples. The thickness over which the composition varies in the growth direction is smaller in the pulse-plated samples obtained in the present work than in any deposit previously reported.

Comparing the composition stabilization thickness of the present d.c.-plated samples with a few earlier results obtained for different sample compositions, it can be seen that the stabilization period during the electrodeposition of various alloys cannot be directly related to other parameters. The stabilization thickness of binary Co–Ni deposits [44] and ternary alloys (Fe–Co–Ni [43] and Cu–Co–Ni [44]) was found to be 150–200 nm. However, the composition stabilization thickness is approximately 90 nm for binary Ni–Cd deposits [44]. By comparing the sample preparation data, it can also be revealed that the concentration of the ions of the minority alloy component in the bath cannot be a determining parameter either.

It is observed for several d.c.-plated binary and ternary alloys [42–44] that the mole fraction of the preferentially deposited minority component has a minimum in the near-substrate transient zone. The occurrence of this minimum is explained with the formation of an unstable depleted solution layer around the cathode and the temporary extension of the diffusion field in the unstirred solution in the early phase of the deposition beyond the distance at which the diffusion field can be stable during the steady-state deposition. Once natural convection begins to play a role and deposition has reached steady-state, the depleted zone shrinks and the concentration gradients near the cathode increase. Hence, the mole fraction minimum of the preferentially deposited component is attributed to a change in the mass transport in the electrolyte as steady-state is achieved. By considering that the mole fractions of the minority components in both Fe–Co–Ni [43] and Cu–Co–Ni [44] ternary alloys are strongly correlated with each other, this explanation seems to be feasible.

The mole fraction minimum of Fe in the present Fe–Ni deposits is occasionally observed, too. The absence of the concentration minimum of the preferentially deposited component (here, Fe) is completely in line with the above assumptions for low duty cycles. In this case, the steady-state depletion zone probably does not extend far enough into the solution to a distance where natural convection plays a significant role. It is also possible that the convective motion of the electrolyte ends by the time when the next current pulse begins. In either of the cases, the temporary extension of the depletion zone is not possible.

In spite of all these uncertainties, the key trend can be seen in Fig. 7. The smaller the duty cycle (or the on-time), the smaller the composition change in the deposit. A minimum of the Fe mole fraction can only be detected if the concentration change in the initial zone is large enough. In Fig. 7, the curves belonging to $\varepsilon = 0.5$ and 0.6 exhibit a minimum before the concentration stabilization.

It must be noted that the reverse composition depth profile analysis has the advantage that the near-substrate zone can be imaged with an unprecedented accuracy. The destructive depth profile analysis methods performed with the conventional direction usually cause a significant signal convolution ("smear out") when the sputtering zone reaches the substrate, which makes it impossible to detect fine composition changes in the near-substrate zone. This may explain why the near-substrate composition change in electrodeposited Fe–Ni samples remained hidden in some earlier studies [52]. Laterally uneven sputtering (i.e., non-ideal crater shape formation) and roughening during the sputtering process may contribute to this signal convolution and make the accurate analysis even more difficult.

The achievement of deposits with even composition along the growth direction is crucial for controlling magnetic properties as a function of the deposit thickness. As clearly shown in the work of Lommel and Girard [38], the resulting coercivity and anisotropy in electrodeposited Fe–Ni samples reflect an interplay of the spontaneous composition modulation along the growth direction and the thickness itself. In order to eliminate the impact of the varying composition, homogeneous samples have to be deposited. The composition gradient formed during d.c. plating may have a significant impact on the residual stress in the deposit, which also leads to an increase in the coercive field of the samples. As shown above, pulse-plating with a sufficiently small duty cycle can open the way to obtain homogeneous Fe–Ni deposits. The main goal of the optimization remains the preparation of electrodeposited Permalloy ($Ni_{80}Fe_{20}$), for which a duty cycle significantly smaller than 0.2 (the usual value in this study) has to be applied.

## 5. Conclusions

Reverse composition depth profiles of d.c.-plated and pulse-plated Fe–Ni samples have been measured with SNMS. The composition of the near-substrate zone of the deposits could be determined with high accuracy. It has been found that the smaller the duty cycle, the smaller the difference in the initial and steady-state composition of the pulse-plated samples The threshold duty cycle at which the initial composition change is negligible becomes smaller as the $Fe^{2+}$ concentration in the electrolyte is reduced. Pulse plating is preferable to d.c. plating to obtain thin deposits ($d < 200$ nm) with uniform composition.

## Acknowledgement

This work was performed with the help of Hungarian Scientific Research Fund (OTKA) through Grant # K 75008.